\documentclass[12pt]{JHEP3}
\usepackage{amsmath}
\usepackage{amssymb}
\usepackage{epsfig}
\usepackage{cite}

\def\beq{\begin{equation}}
\def\eeq{\end{equation}}
\def\half{{\textstyle {\frac12}}}
\def\sfrac#1#2{{\textstyle {\frac{#1}{#2}}}}
\def\cG{{\cal G}}
\def\cN{{\cal N}}
\def\cP{{\cal P}}

\def\om{\omega}
\def\cO#1{{\cal{O}}\left(#1\right)}
\def\hW{\widehat{W}}
\def\abs#1{\left|#1\right|}
\def\ab{\bar{\alpha}_s}

\def\beeq{\begin{eqnarray}}
\def\eeeq{\end{eqnarray}}
\def\bsub{\begin{subequations}}
\def\esub{\end{subequations}}

\def \as{\relax\ifmmode\alpha_s\else{$\alpha_s${ }}\fi}
\def \asb{\relax\ifmmode\bar{\alpha}_s\else{$\bar{\alpha}_s${ }}\fi}
\def \al{\relax\ifmmode\alpha\else{$\alpha${ }}\fi}
\def\lrang#1{\left\langle#1\right\rangle}

\def\vec#1{{\bf #1}}

\date{}
\title{Monte Carlo and large angle gluon radiation} 

\author{Yu.L.\ Dokshitzer$^{1,3}$\footnote{On leave of absence: St.\
Petersburg Nuclear Physics Institute, 188350, Gatchina, Russia} $\>$
and G.~Marchesini$^{2}$ \\ \normalsize $^1$LPTHE, Universities of
Paris-VI and VII and CNRS, Paris, France\\ \normalsize $^2$University
of Milan--Bicocca and INFN Sezione di Milano--Bicocca, Milan, Italy \\
\normalsize $^3$CERN Theory Division, Geneva, Switzerland}

\abstract{We discuss the problem of incorporating recoil effects into
  the probabilistic QCD evolution scheme based on the picture of
  colour dipoles as done in recent Monte Carlo programs. Such a scheme
  correctly describes subleading soft contributions to multiplicity
  distributions.  However we find that a {\it simple}\/ receipt for
  incorporating recoil effects into the dipole multiplication picture
  based on the ordering of gluon {\em energies}\/ conflicts the
  collinear factorization and does not lead to the correct DGLAP
  equation. }

\begin{document}

\maketitle


\section*{Preface}

We share opinion of our late friend Bo Anderson who once bitterly
complained about two aspects of scientific publishing that became
standard in recent decades: 1) shying away from expressing emotions
and 2) not making public an outcome of a study that produced a {\em
  negative}\/, undesired results.  The present paper is moderately
emotional; at the same time it is perfectly in line with the second
Bo's demand: it reports the finding that the authors did not plan, and
would rather not want, to arrive at.

The answer to the question that is posed in this paper suggests that a
formulation of Monte Carlo event generation based on the
multiplication picture of energy ordered colour dipoles is plugged by
a serious problem.
  
\section{Introduction}

Generation of events using Monte Carlo methods is an indispensable
tool for planning, running and analysing the results of modern high
energy experiments.  A possibility to generate multi-particle
production as a Markov chain of successive independent parton
splittings is based on the general property of {\em factorization of
  collinear singularities}\/ which is characteristic for logarithmic
field theories (quantum field theories with dimensionless coupling).
 
Event generation is necessarily an approximate procedure.  Strictly
speaking, in order to predict the cross section $d\sigma_n$ for
production of $n$ partons in a hard interaction with the hardness
scale $Q^2$, one needs to plug in the QCD matrix element of the (at
least) $n^{\mbox{\scriptsize th}}$ order of the perturbation theory.
Instead, one repeats $n$ times the basic $1\to2$ splitting process of
the first order in $\as$. By so doing one correctly samples the {\em
  major part}\/ of the $n$-parton phase space, though not all of it.
It is this part that gives the dominant, maximally collinear enhanced
contribution to, e.g., the inclusive particle distribution,
$D^{(n)}=\cO{\as^n \log^n Q^2}$.  This approximation known as {\em
  leading logarithmic}\/ (LLA) can be systematically improved by
including higher order parton splitting processes.  Thus, the
next-to-leading logarithmic accuracy corresponding to the series
$\as^{n+1}\log^n Q^2$ is achieved by incorporating $1\to 3$ parton
splittings, etc.  For inclusive parton distributions this leads to the
DGLAP evolution equations \cite{DGLAP} whose generalisation to
multi-particle distributions can be achieved in the spirit of
jet-calculus \cite{KUV}, with additional account of soft gluon
coherence.

Beyond the one loop ``evolution Hamiltonian'' it starts to matter how
one organises the parton chain, that is, what one chooses as ``parton
evolution time''.  In particular, due to the presence of {\em soft
  gluons}\/ one encounters corrections of the type $(\as\log^2 x)^k$
which are formally subleading (non-collinear) but become explosively
large in the $x\ll 1$ limit (here $x$ is the gluon energy fraction).
This eventuality can be cured by an appropriate tuning of the
evolution time parameter.  Namely, for time-like parton multiplication
(jets) one has to choose the {\em angular}\/ ordering \cite{AO}.  This
choice corrects the ``naive'' {\em fluctuation time}\/ ordering
(dictated by an examination of Feynman denominators in a cascading
process) by taking into clever account collective effects leading to
destructive interference in soft gluon radiation (for more details
see, e.g., \cite{BCM,EAO,DT,A,Basics}).

An alternative way of dealing with soft gluon interference effects for
time-like parton multiplication (jets) is provided by the ``dipole
scheme'' \cite{BCM,Gust} in which an independently radiating {\em
  parton}, $1\to 2$, is replaced by a colourless {\em dipole}\/ formed
by two partons neighbouring in the colour space, $2\to 3$.  Gluon
radiation off a dipole is automatically suppressed at angles exceeding
the dipole opening angle thus reproducing the angular ordering.
Colour suppressed contributions $\cO{1/N_c^2}$ that lie beyond the
scope of the dipole approach are numerically small and difficult to
trigger experimentally~\cite{AzAss}.

Within the standard coherent parton cascade picture
\cite{EAO,Basics} it is the small-angle multiplication
processes populating jets that enjoy full all-order treatment (get
``exponentiated'').  The dipole formulation offers a possibility to
improve this treatment by taking into consideration logarithmically
enhanced effects due to multiple emission of soft gluons at {\em large
  angles}\/ with respect to jets.  Non-collinear soft gluons dominate
inter-jet particles flows in various hard processes.  They also
complicate the analysis of the so-called {\em nonglobal}\/ QCD
observables \cite{DasSal,ng-plus,DM_LargeAng}, i.e. in observables in
which recorded radiation is confined in geometrically definite phase
space regions.
It is then interesting to involve these corrections into a Monte
Carlo, code based on dipole emission, including soft radiation awy
from jets.

In this paper we discuss a dipole based scheme (in the large-$N_c$
approximation) which is well suited for deriving improved analytic
predictions for observables like mean multiplicities, inclusive soft
particle spectra and energy flows, correlations, etc., that
incorporate large-angle soft gluon radiation effects as in
\cite{DasSal,ng-plus,DM_LargeAng}. At the same time, we observe that
once one aims at {\em beyond the no-recoil (soft) approximation},
treating colour dipoles as independently evolving entities is likely
to conflict the collinear factorization. It does not lead to the
correct DGLAP equation.

\medskip 
The paper is organised as follows.  We start by constructing
in Section 2 an evolution equation for the generating functional that
describes soft gluon content of $e^+e^-$ annihilation events.  No
recoil is included. This equation is based on the factorization of the
multi-soft gluon distribution, in the planar limit, and uses the
centre-of-mass energy as an ``evolution time''.  Contrary to what is
done normally, no angular variable enters the evolution time here.
The evolution equation in the integral form we interpret as a Markov
process of successive dipole branchings, which interpretation leads to
a Monte Carlo chain process for probabilistic event generation.  As a
first example, we solve the evolution equation for mean gluon
multiplicity to analyse subleading corrections due to large-angle soft
gluon radiation.  Here we work in the {\em soft approximation}\/
meaning: 1) small energies of emitted gluons and 2) neglecting the
energy-momentum loss (recoil) by the primaries.

These two aspects of the ``soft gluon approximation'' do not
necessarily go together.  In Section~3 we consider an inclusive energy
distribution of the leading quark in the quasi-elastic limit,
$(1-x)\ll 1$, where emitted gluons ought to be soft but the quark
recoil is essential.  At the beginning, in the construction of the
evolution equation no recoil effects were included. They are
incorporated later by using the recoil strategy that seems natural for
a dipole multiplication scheme, see \cite{CatSey}.  We find, however,
that the quark fragmentation function so obtained deviates essentially
from the collinear resummation result given by the DGLAP evolution
equation.  This is in contrast to what is happening in Monte Carlo
schemes based on angular ordered time-like cascades where collinear
singularities are correctly resummed \cite{EAO,HERWIG}.

Discussion of the problem encountered is relegated to the Conclusions
section.

\section{Multiple soft gluons and Monte Carlo}

The method of generating functionals makes it straightforward to
generate exclusive events using Monte Carlo techniques.  We start from
the known $n$-gluon distribution in the soft limit and construct the
evolution equation for the corresponding generating functional.

The generating functional for a quark-antiquark pair $q\bar{q}$ plus
an ensemble of $n$ secondary partons, $\gamma^*\to p_ap_b\,
q_1,q_2,\ldots q_n\,, $ is
\begin{equation}\label{eq:Gab}
{G}(Q,u) = \sum_{n=0}^\infty \int d\Phi^{(n)}_{ab}(q_1,q_2,\ldots,q_n) \,
u_au_b \prod_{i=1}^n u(q_i) \abs{M^{(n)}_{ab}(q_1,\ldots,q_n)}^2 .
\end{equation}
Here $p_a$, $p_b$ are the final on-mass-shell momenta of the quark and
antiquark produced by the virtual photon $\gamma^*$ and $q_i$ the
momenta of the secondaries.  The matrix element $M^{(n)}_{ab}$
describes production of $n+2$ real final state partons and contains
any number of virtual ones.  $d\Phi^{(n)}$ is the full phase space
factor of the ensemble of $n+2$ massless particles:
\begin{equation}
  d\Phi^{(n)}_{ab}(q_1,q_2,\ldots,q_n) = (dp_a) (dp_b) \prod_{i=1}^n
  (dq_i) \> (2\pi)^4\delta^{(4)}(p_a+p_b+\sum_{i=1}^n q_i-Q),
\end{equation}
where 
\[ (dq)\>=\> \frac{d^3 \vec{q}}{2\omega\,(2\pi)^3}, \quad
 \omega=q_{0}= \abs{\vec{q}}.
\]
The ``source functions'' $u$ are attached to each parton in
\eqref{eq:Gab}. They help to extract an arbitrary final state
observable once the functional is known.  The fully inclusive
measurement, that is when one allows for production of any number of
particles with arbitrary momenta, corresponds to setting all
$u\!=\!1$. This gives $G(Q,u\!=\!1)=1$ corresponding to normalization
to the total cross section.  In this section we work in the no-recoil
approximation.

\subsection{Generating functional in the soft limit}

In the large-$N_c$ (``planar'') approximation the squared matrix
element can be approximated at tree level as
\begin{subequations}\label{eq:BCM-ant}
\begin{equation}
\abs{M^{(n)}_{ab}(q_1,\ldots,q_n)}^2 \>\simeq \> \frac{(4\pi^2\ab)^n}{n!}
\sum_{\mbox{\scriptsize{perm}}} W_{ab}(q_{i_1},\ldots,q_{i_n}) 
\cdot \abs{M_{ab}^{(0)}}^2\,\qquad \,\ab=\frac{N_c\as}{\pi},
\end{equation}
where $W_{ab}$ is the ``multiple antenna'' function \cite{BCM}
describing the production probability of $n$ soft real gluons $q_i$:
\begin{equation}\label{eq:mult-antenna}
 W_{ab}(q_1,q_2,\ldots,q_n) \>=\>
 \frac{(p_ap_b)}{(p_aq_1)(q_1q_2)\cdots (q_np_b)} .
\end{equation}
\end{subequations}
This result holds in the {\em soft gluon approximation}\/ $q_i\ll
p_a,p_b$. In invariant terms,
\begin{equation}
(q_ip_a),\, (q_ip_b) \>\ll\> (p_ap_b), 
\end{equation}
or, equivalently, $(q_i,p_a+p_b) \>\ll\> 2(p_ap_b)=Q^2$ giving
\begin{equation}
\label{eq:energy}
\omega_i \>\ll\> E_a\simeq
  E_b \simeq Q/2,
\end{equation}
with $E_a,E_b$ and $\omega_i$ the energies in the centre-of-mass of
the ``dipole'' $(ab)$. In this approximation the phase space can be
simplified as
\begin{equation}
d \Phi^{(n)}_{ab}(q_1,\ldots,q_n)\> \simeq\> (dp_a) (dp_b)
(2\pi)^4\delta^4(p_a+p_b-Q) \cdot \prod_{i=1}^n (dq_i) \>\equiv\>
d\Phi_{ab}^{(0)}\cdot \prod_{i=1}^n (dq_i)\,.
\end{equation}
The expression \eqref{eq:BCM-ant} was derived implying {\em strong
  ordering}\/ of gluon 
  energies:
\[
\omega_{i_n} \ll \omega_{i_{n-1}} \ll \ldots \ll \omega_{i_2} \ll
\omega_{i_1} \ll Q/2.
\]
The answer, however, is symmetric with respect to momenta of
participating gluons.  Therefore, we may ignore the ordering between
gluons and set a common upper bound $E=\half Q$ for gluon energies.

Extracting the quarks whose momenta in the soft gluon approximation
are not affected by radiation, we may write
\begin{equation}
   G(u;Q) \simeq \int d\Phi_{ab}^{(0)}\, u(p_a)u(p_b)\,|M_{ab}^{(0)}|^2 \cdot
\cG(p_a,p_b; \half Q).
\end{equation}
The functional $\cG(p_a,p_b;E)$ describes production of any number of
gluons with energies limited from above by some value $E$, off the
dipole formed by the quarks with momenta $p_a$, $p_b$.  By setting
$E=\half Q$ we obtain the multi-gluon generating functional describing
$e^+e^-$ annihilation process with $s=Q^2$.

The generating functional contains proper virtual factors for each
$n$-gluon contribution. Omitting for the moment the virtual
corrections, we have
\begin{equation}\label{eq:GF-soft-E} 
\cG_{(\rm real)}(p_a,p_b; E) \equiv \sum_n \int W_{ab}(q_1,\ldots,q_n)\,
\prod_{i=1}^n \bigg[ 
\omega_i d\omega_i\frac{d\Omega}{4\pi}\,\ab 
\cdot u(q_i) \cdot
\vartheta(E\!-\!\omega_i) \bigg] .
\end{equation}
Due to the symmetric structure of the antenna function
\eqref{eq:BCM-ant} one may choose $W_{ab}(q_1\ldots q_n)$ to represent
any {\em energy ordered}\/ ensemble of
gluons 
and drop the symmetry factor $1/n!$.

\subsection{Evolution equation}

To obtain an evolution equation for the generating functional we need
to exploit the structure of the multi-gluon antenna distribution
\eqref{eq:BCM-ant} and to deduce a recurrence relation. To construct a
recurrence relation one should select an ``evolution variable'' to
order emitted gluons.  In the logic of the {\em collinear
  approximation}\/ this is done by ordering the angles of successive
gluon emissions, which ordering takes full care of the destructive
interference contributions in the soft region \cite{AO,BCM,EAO,Basics}
and preserves the probabilistic parton multiplication picture.  In
particular, this procedure was applied to the expression
\eqref{eq:GF-soft-E} to construct the HERWIG event generator
\cite{EAO,HERWIG}.

However, as was already stated, the distribution \eqref{eq:BCM-ant} is
valid in the soft limit for {\em arbitrary angles}\/, and one is
tempted to use this property in order to lift off the collinear
approximation.  The derivation of the evolution equation avoiding the
small angle approximation was done in
\cite{DasSal,ng-plus,DM_LargeAng} where the upper bound $E$ on gluon
energies in the event centre-of-mass was treated as the ``evolution
time'' parameter.  Let us recall the corresponding construction
employing the energy as an evolution variable.

\subsubsection{Energy as an evolution variable} 
We take the ordering in the energy of emitted gluons in the
center-of-mass of the event \eqref{eq:energy} and deduce an evolution
equation for the soft generating functional.
The evolution equation follows from the exact recurrence relation ---
the {\em factorization property}\/ of the multi-gluon antenna function
\eqref{eq:mult-antenna} --- which reads
\begin{equation}
W_{ab}(q_1,q_2,\ldots,q_n) = W_{ab}(q_\ell) \cdot W_{a\ell}(q_1,\ldots
q_{\ell-1}) \, W_{\ell b}(q_{\ell+1},\ldots, q_n),
\end{equation}
where $\ell$ is any gluon.  To construct the evolution equation one
takes $\ell$ to be the {\em most energetic}\/ among the gluons.

Introducing the scaled antenna function
\[
 \hW_{ab}(q) \equiv \omega^2 \cdot W_{ab}(q)\>=\>
 \frac{\xi_{ab}}{\xi_{aq}\,\xi_{bq}}\,, \qquad \xi_{ik} =
 1-\cos\Theta_{ik}\,,
\]
which depends only on the angles between partons, we obtain
\begin{equation}\label{eq:softGF}
 E\partial_E \cG(p_a,p_b;E) = \int \frac{d\Omega}{4\pi} \ab\,
\hW_{ab}(q) \bigg[ u(q)\, \cG(p_a,q;\omega)\cG(q,p_b;\omega) -
\cG(p_a,p_b,\omega)\bigg]_{\omega=E}\!\!\!\!\!\!\!\!.
\end{equation}  
Here the subtraction term takes care of virtual corrections, so that
taking all $u(q)=1$ one derive the desired normalization
\begin{equation}
\label{eq:unity}
\left. \cG(p_a,p_b;E)\right|_{u=1}=1\,.
\end{equation}
No recoil was included up to now; it will be considered later.

\subsubsection{Integral evolution equation and Monte Carlo}
By exponentiating the total one-gluon emission probability we obtain
the {\em Sudakov form factor}\/ which describes the probability that
the dipole does not radiate gluons with energies up to a given value
$E$:
\begin{equation}
 S(p_a,p_b;E) = \exp\left\{ -\int^E 
\frac{d\omega}{\omega}\frac{d\Omega}{4\pi}\,
\asb(q^{ab}_t)\, \hW_{ab}(q)
 \, \vartheta(q_t^{ab} - Q_0) \right\}.
\end{equation}
Here $q_t^{ab}$ is the invariant transverse momentum 
of the gluon with respect to the pair of quarks: 
\begin{equation}
  (q_t^{ab})^2 = \frac{2(p_aq)(p_bq)}{(p_ap_b)}\>=\>
  \frac2{W_{ab}(q)}\,,
\end{equation}
and an arbitrary parameter $Q_0$ has been introduced as a collinear
cutoff.  By constructing the logarithmic derivative of the Sudakov
form factor over the maximal gluon energy,
\begin{equation}
\left.  E\partial_E \ln S(p_a,p_b;E) = - \int \frac{d\Omega}{4\pi}
  \ab(q_t^{ab})\, \hW_{ab}(q)\, \vartheta(q_t^{ab} - Q_0)\right|_{\omega=E},
\end{equation}
and plugging this expression into \eqref{eq:softGF} we can trade the
virtual subtraction term for the Sudakov factor.
By so doing we arrive at the equivalent equation for the soft
generating functional in the integral form:
\begin{equation}\label{eq:soft-GF-int}
\begin{split}
  \cG(p_a,p_b;E) \>=& \>\> 
S(p_a,p_b;E) + \int^E
\frac{d\omega}{\omega}\int \frac{d\Omega}{4\pi}\, \,u(q)\cdot
\frac{S(p_a,p_b;E)}{S(p_a,p_b;\omega)} \\ & \cdot \ab(q_t^{ab})\,
\hW_{ab}(q)\, \vartheta(q_t^{ab} - Q_0) \cdot \cG(p_a,q;\omega)\,
\cG(q,p_b;\omega) .
\end{split}
\end{equation}
Iteration of this equation can be interpreted as parton branching
which can be realised as a Monte Carlo (Markov) process.  To see this,
observe that the iteration kernel
\begin{equation}\label{eq:soft-dP}
\begin{split}
d\cP(p_a,p_b,q;E)\>=\> \frac{d\omega}{\omega}\frac{d\Omega}{4\pi}\,
\frac{S(p_a,p_b;E)}{S(p_a,p_b;\omega)} \cdot \ab(q_t^{ab})\,
\hW_{ab}(q)\, \vartheta(q_t^{ab} - Q_0)
\end{split}
\end{equation}
can be written as the product of two probability distributions (here
the boundary $q_t^{ab}\,>\,Q_0$ is implicit):
\begin{equation}\label{eq:soft-drdR}
\begin{split}
&d\cP(p_a,p_b,q;E)\>=\> d\cP^{(1)}_{ab}(\omega,E)\cdot
d\cP_{ab}^{(2)}(\Omega_q)\,,\\ &\cP^{(1)}_{ab}(\omega,E)=
\frac{S(p_a,p_b;E)}{S(p_a,p_b;\omega)}\,,\qquad
d\cP_{ab}^{(2)}(\Omega_q)=
\ab\,\frac{d\Omega_q}{4\pi}\,\frac{\xi_{ab}}{\xi_{aq}\xi_{qb}}
\left(\int\ab\,\frac{d\Omega}{4\pi}\,
\frac{\xi_{ab}}{\xi_{aq}\xi_{qb}}\right)^{-1}.
\end{split}
\end{equation}
In the Monte Carlo process one tries to generate emission of a gluon
with momentum $q$ off a given dipole.  The first distribution,
$d\cP^{(1)}$, provides the gluon energy $\omega$, if the transverse
momentum bound is satisfied, and the second one, $d\cP^{(2)}$, its
direction $\Omega_q$.  If the boundary is not satisfied, the dipole
does not emit.  One has
\begin{equation}\label{eq:no-emission}
\int d\cP(p_a,p_b,q,E)\>=\>1-S(p_a,p_b;E),
\end{equation}
which shows that the Sudakov factor gives the probability of not
emitting a soft gluon within the resolution $q_t^{ab}\,>\,Q_0$.

\subsection{Mean multiplicity}

By construction the generating functional \eqref{eq:soft-GF-int}
embeds only soft gluon radiation. Incorporating double logarithmic
(simultaneously soft- and collinear-enhanced) contributions, it also
contains single logarithmic effects due to emission of soft gluons at
{\em large angles}.

Here we demonstrate an application of the dipole-based evolution to
the calculation of subleading large-angle soft gluon corrections. In
the collinear approach, such corrections are treated as due to
``multi-jet'' configurations (contributing to the ``coefficient
function'') rather than jet evolution (``anomalous dimension''), see
\cite{Basics}.

Consider mean gluon multiplicity.  
In order to obtain an equation for the multiplicity of secondary partons, one
applies to \eqref{eq:softGF} the variational derivative over the probing function $u(k)$ and integrates over $k$, while setting $u\equiv 1$ for all remaining probing functions (one-particle inclusive measurement): 
\begin{equation}
E\partial_E N(\xi_{ab},E) = \int \frac{d\Omega}{4\pi}\,
\ab(k_t^{ab})\,\hW_{ab}(q) \bigg[ 1+ N(\xi_{qa},E) + N(\xi_{qb},E) -
N(\xi_{ab},E) \bigg]_{\omega=E} .
\end{equation}
Introducing $\cN = 1+ N$ we have \cite{MM} 
\begin{equation}\label{eq:NdiffDL}
\cN'(\xi_{ab})=
E\partial_E \cN(\xi_{ab}) = \int \frac{d\Omega}{4\pi}\,
\ab(k_t^{ab})\,\hW_{ab}(q) \bigg[ \cN(\xi_{qa}) + \cN(\xi_{qb}) -
\cN(\xi_{ab}) \bigg] .
\end{equation} 
In order to illustrate the difference with the standard approach based
on the angular ordering we will analyse \eqref{eq:NdiffDL} in the
double logarithmic approximation which neglects hard parton splittings
and recoil effects.  For the time being we will ignore the running of
the coupling; the corresponding corrections will be addressed later.

Representing the integrand on the r.h.s.\ of \eqref{eq:NdiffDL} as
\[
  \frac{\xi}{\xi_{1}\xi_2} \bigg[ \cN(\xi_{1}) + \cN(\xi_{2}) -
\cN(\xi) \bigg] = \frac{1}{\xi_2}\left[
\frac{\xi}{\xi_1}\cN(\xi_1)-\cN(\xi)\right] + \frac{1}{\xi_1}\left[
\frac{\xi}{\xi_2}\cN(\xi_2)-\cN(\xi)\right] -
\frac{\xi-\xi_{1}-\xi_{2}}{\xi_{1}\xi_{2}}\cN(\xi),
\]
and using the $1\leftrightarrow 2$ symmetry, we obtain
\begin{equation}
\cN'(\xi) =  \int \frac{d\Omega}{4\pi}\, \ab \left\{ 
 \frac{2}{\xi_2}\left[ \frac{\xi}{\xi_1}\cN(\xi_1)-\cN(\xi)\right]
-  \frac{\xi-\xi_{1}-\xi_{2}}{\xi_{1}\xi_{2}}\cN(\xi)\right\} .
\end{equation}
We can perform the integration over the azimuth of $\vec{q}$ around
the direction of $\vec{p}_a$,
\[
 \int \frac{d\Omega}{4\pi} = \frac12 \int_0^2 d\xi_1 \cdot \int
 \frac{d\phi_1}{2\pi}.
\]
Using the relation
\[
\lrang{\frac1{\xi_2}} \equiv \int\frac{d\phi_1}{2\pi} \frac1{\xi_2} =
\frac1{\abs{\xi-\xi_1}},
 \]
we get
\begin{equation}
\lrang{\frac{\xi-\xi_1-\xi_2}{\xi_1\xi_2}} =
 \frac1{\xi_1}\,\left(\frac{\xi-\xi_1}{\abs{\xi-\xi_1}} -1\right) =
 -\frac{2}{\xi_1}\, \vartheta(\xi_1-\xi).
\end{equation}
As a result, 
\begin{equation}\label{eq:NdiffDL-2}
  \cN'(\xi) = \ab \int_0^{2} \frac{d\xi_1}{\abs{\xi-\xi_1}} 
  \left[ \frac{\xi}{\xi_1} \cN(\xi_1) - \cN(\xi)\right]
+ \ab\int_\xi^2 \frac{d\xi_1}{\xi_1}\, \cN(\xi),
\end{equation}
with the $E$ dependence implicit.
The first contribution we split into integrals over small and large
angles, $0\le \xi_1\le \xi$ and $\xi<\xi_1\le2$.  Introducing the
integration variable $\eta=\xi_1/\xi\le 1$ in the first region and
$\eta=\xi/\xi_1$ in the second, $\half\xi \le \eta\le 1$, the equation
\eqref{eq:NdiffDL-2} takes the form \cite{MM}
\beeq
\cN'(\xi) &=&\int_0^{1} \frac{d\eta\> \ab}{1-\eta} \left[ \frac1\eta
\cN(\eta\xi) - \cN(\xi)\right] +\int_{\half\xi}^1 \frac{d\eta\>
\ab}{1-\eta} \left[ \cN\big(\eta^{-1}\xi\big) - \cN(\xi)\right] .
\eeeq 
This result we represent as a sum of two terms,
\begin{subequations}
\begin{equation}\label{eq:cN2}
E\partial_E \cN(\xi) = \int_0^{1} \frac{d\eta\> \ab}{\eta}
\cN(\eta\,\xi) \>+\> \Delta(\xi),
\end{equation}
where
\begin{equation}\label{eq:DcN}
\Delta(\xi) = \int_0^{1} \frac{d\eta\> \ab}{1-\eta} \big[
\cN(\eta\,\xi) - \cN(\xi)\big] \>+\>\int_{\half\xi}^1
\frac{d\eta\> \ab}{1-\eta} \big[ \cN\big(\eta^{-1}{\xi}\big) -
\cN(\xi)\big] . \quad { }
\end{equation}
\end{subequations}
The first integral term on the r.h.s.\ of \eqref{eq:cN2} generates the
standard DL anomalous dimension for the mean multiplicity; the second
one constitutes a subleading correction.

\subsubsection{The main term}
We have to solve  the equation 
\begin{equation}\label{eq:cNDL}
  E\partial_E \cN^{\mbox{\scriptsize DL}}(\xi;E) = 
\int_0^{1} \frac{d\eta\> \ab}{\eta}\> 
\cN^{\mbox{\scriptsize DL}}(\eta\,\xi;E)
\end{equation}
with the initial condition
\[
\left.   \cN(\xi; E)\right|_{E\sqrt{2\xi}=Q_0}\>=\> 1. 
\]
The solution is a function of a single variable $Q_\xi \equiv
E\sqrt{2\xi}$ (the maximal transverse momentum of partons radiated by
the dipole).

The argument of the running coupling in \eqref{eq:cNDL} can be
approximated as
\[
 \ab(k_t^{ab}) = \ab \bigg( E\sqrt{\frac{2\xi_1\xi_2}{\xi}}\bigg)
\simeq \ab \bigg( E\sqrt{{2\xi_1}}\bigg) .
\]
The correction to this approximation turns out to be negligible since
the logarithmic factor $\ln(\xi_2/\xi)$ vanishes in the collinear
limit $\xi_1\to 0$:
\[
   \propto \int \frac{d\xi_1}{\xi_1} \, \big(\beta_0\ab^2\big) \cdot
   \ln\frac{\xi_2}{\xi} \>=\>\cO{\ab^2}.
\]
Thus we obtain the double differential equation
\begin{equation}\label{eq:dd}
 \left(\frac{d^2}{d\ln Q_\xi}\right)^2 \cN^{\mbox{\scriptsize DL}}(Q_\xi) 
\>=\> 2  \ab(Q_\xi)\, \cN^{\mbox{\scriptsize DL}}(Q_\xi), \qquad 
Q_\xi = E\sqrt{2\xi},  
\end{equation}
with the initial conditions $\cN(Q_0)=1$ and $\cN'(Q_0)=0$. 
Its solution for the fixed coupling case reads
\begin{equation}\label{eq:cNexp}
  \cN^{\mbox{\scriptsize DL}}(Q_\xi) = \cosh\left(\gamma\ln
 \frac{Q_\xi}{Q_0}\right)
   \simeq \frac12\left(\frac{Q_\xi}{Q_0}\right)^\gamma,
\end{equation}
where the {\em anomalous dimension}\/ $\gamma$ is given the correct
expression
\[
 \gamma = \ab \int_0^1 \frac{d\eta}{\eta} \cdot \eta^{\half \gamma}\>
 =\> \ab \cdot \frac{2}{\gamma} \quad \Longrightarrow\quad \gamma =
 \sqrt{2\ab}\,.
\]

\subsubsection{The correction term}
Let us analyse the correction term \eqref{eq:DcN} keeping the
contributions of the order of $\ab^{3/2}$ while neglecting
contributions $\cO{\ab^2}$.  The answer depends on the value of the
opening angle~$\xi$.

\paragraph{Small opening angles ($\xi\ll 1$).}

It is the region of small opening angles that gives a dominant
contribution to the integral determining the DL anomalous dimension in
\eqref{eq:cNDL}.  In this entire region the correction \eqref{eq:DcN}
does not contribute at the $\ab^{3/2}$ level.  Indeed, substituting
the solution \eqref{eq:cNexp} into the expression \eqref{eq:DcN} and
setting $\xi\to 0$ in the lower limit of the second integral we get
\begin{equation}
\begin{split}
 \frac{ \Delta(\xi;E)}{ \cN(\xi;E)} =&\> \ab \int_0^1\frac{d\eta}{1-\eta}
  \left[ \eta^{\gamma/2} + \eta^{-\gamma/2} -2 \right] \\ =& \> \ab
  \cdot \bigg[ 2\psi(1)-\psi(1+\half\gamma) -\psi(1-\half\gamma)
  \bigg] = \ab\cdot \sfrac{1}{4}\gamma^2\,\psi''(1) + \ldots \simeq \half
  \ab^2\,\psi''(1).
\end{split}
\end{equation}

\paragraph{Large opening angle ($\xi\simeq 2$).}
At the same time, $\Delta$ contributes to the multiplicity of the
``fully open'' dipole, $\xi=2$. The expression \eqref{eq:DcN} for
$\Delta(2;E)$ contains one integral:
\[
 \Delta(2;E) = \ab \int_0^1\frac{d\eta}{1-\eta}\big[\,\cN(\eta\cdot
 2;E) - \cN(2;E)\,\big].
\]
Expanding the difference of the multiplicity factors,
\[
 \cN(\eta\cdot 2;E) - \cN(2;E) \>=\> E\partial_E \cN(2;E)\cdot
 \half\ln\eta + \ldots,
\]
we derive
\[
 \Delta(2;E) \simeq E\partial_E \cN(2;E)\cdot \ab\,
 \int_0^1\frac{d\eta}{1-\eta}\cdot \half \ln\eta =
 -\frac{\pi^2}{12}\ab\cdot E\partial_E \cN(2;E).
\]
Integrating this correction over energy we finally obtain an order
$\as$ correction to the dipole multiplicity due to emission of {\em
  two}\/ soft energy ordered gluons:
\begin{equation}
  \cN(2;E) = \cN^{\mbox{\scriptsize DL}}(2;E)\cdot
  \bigg(1-\frac{\pi^2}{12}\ab + \ldots\bigg).
\end{equation}
In the standard approach, this correction belongs to the {\em
  coefficient function}\/ and originates from a ``four-jet''
configuration with two large-angle energy ordered soft gluons as
additional ``jets'' \cite{Basics}.

\subsubsection{Other observables sensitive to large-angle gluon effects}

There is a variety of observables that are sensitive to soft gluon
radiation at large angles.  Among them one can mention production of
heavy quark pairs in jets \cite{MM}, non-global jet observables
\cite{DasSal} including various particle distributions and
correlations related with particle flows in the inter-jet regions
\cite{ng-plus,DM_LargeAng,BMS,string,MWassociate,S-interjet,DKMW_corr}.
The dipole-based evolution equation \eqref{eq:soft-GF-int} is well
suited for taking into account single logarithmic correction effects
due to multiple soft gluons in these observables and elsewhere.

At the same time, it remains insufficient for building a realistic
Monte Carlo event generator which is impossible without
incorporating full parton decay probabilities, including hard parton
splittings, and of the recoil effects to ensure energy--momentum
conservation.

\section{Attempt to include recoil in dipole multiplication\label{Sec3}}

Generally speaking, a gluon emitted by a dipole cannot be ascribed as
offspring to either of the two partons that form the dipole. However,
in the collinear limit when the gluon momentum $q$ is quasi-parallel
to one of the two hard partons, say, $p_a$, it is the parton $a$ that
can be said to ``independently split'' into two, sharing its momentum,
$P_a\to p_a+q$. A small change in the ``spectator'' momentum $P_b\to
p_b$ which is necessary to compensate for the virtuality of the
$(a,q)$ pair, vanishes in the collinear limit. A general recoil
strategy has to be formulated in such a way that in the collinear
limit the answer reduces to the standard DGLAP parton splitting
function $P_a\to p_a+q$ that includes hard momentum configurations,
$p_a\sim q$. This is a source of a significant single-logarithmic
correction, this time not from soft gluon radiation at large angles
but from hard collinear-enhanced emission.

Thus, in order to properly formulate a recoil strategy, one must split
the soft dipole radiation function $W_{ab}(q)$ into two pieces,
$W^{(a)}_{ab}(q)$ and $W^{(b)}_{ab}(q)$, each of which
incorporates the collinear singularity when $\vec{q}\,||\,\vec{p}_a$
or $\vec{q}\,||\,\vec{p}_b$, respectively:
\begin{subequations}\label{eq:Wsplit}
\begin{equation}\label{eq:Wsplit1}
 W_{ab}(q) = \frac{(p_ap_b)}{(p_aq)(qp_b)} =
  W^{(a)}_{ab}(q) + W^{(b)}_{ab}(q),
\end{equation}
or, graphically,
\begin{equation}\label{eq:Wsplit2}
 \epsfig{file=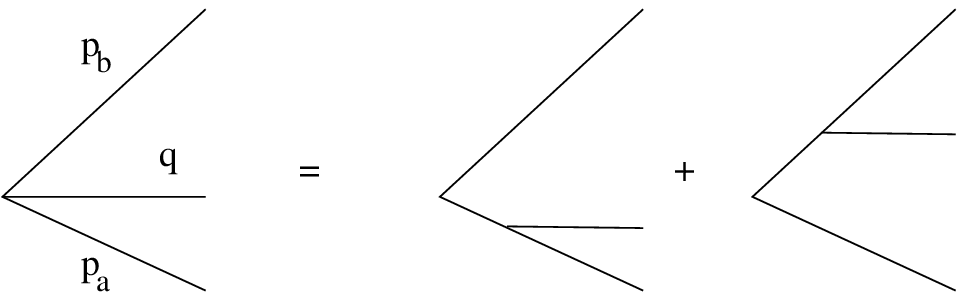}
\end{equation}
\end{subequations}
One can do this differently, for example by introducing \cite{CatSey}
\begin{equation}\label{eq:WCS}
  W^{(a)}_{ab}(q) = \frac{(p_ap_b)}{(p_aq)(p_a+p_b,q)}\,,\quad
  W^{(b)}_{ab}(q) = \frac{(p_ap_b)}{(p_bq)(p_a+p_b,q)}\,.
\end{equation}
Another possibility to split the dipole is given by the ``conditional
probabilities'' \cite{EAO,DT}
\begin{equation}\label{eq:WAO}
  W^{(a)}_{ab}(q) = W_{ab}(q)\cdot
  \half \left[\, 1 + \frac{(qp_a)(p_bQ) -
  (qp_b)(p_aQ)}{(p_ap_b)(qQ)}\,\right] =  
\frac{1}{2\omega^2\,\xi_{qa}} \left[\, 1 +
  \frac{\xi_{ab}-\xi_{qa}}{\xi_{qb}} \,\right].
\end{equation}
The distribution \eqref{eq:WAO} produces the {\em exact angular
ordering}\/ upon averaging over the azimuthal angle $\phi$ of the
gluon momentum $\vec{q}$ around the singular direction $\vec{p}_a$:
\begin{equation}
\label{eq:ang-ord}
\lrang{W_{ab}^{(a)}}_{\phi_{\vec{q},\vec{p}_a}} =
\frac1{\omega^2\xi_{qa}}\cdot
\vartheta\big(\xi_{ab}-\xi_{aq}\big).
\end{equation}
Note that in all cases the distributions $W^{(c)}_{ab}(q)$ ($c=a,b$)
fall fast and essentially become irrelevant when the emission angle
exceeds the opening angle of the parent dipole, $\xi_{qc}>\xi_{ab}$,
that is away from the angular ordered kinematics.

\subsection{Recoil strategy}
Consider the elementary process
\[
    P_a+P_b \>\Longrightarrow\> p_a+p_b+q\,.
\]
One ascribes definite recoil pattern separately to the two terms in \eqref{eq:Wsplit}.  Thus, for the first term $W_{ab}^{(a)}$ which is collinear singular in the direction
$\vec{q}\,||\,\vec{p}_a$, we choose, following the Catani--Seymour prescription \cite{CatSey}
\[
 p^{(a)}_b = (1-y)P_b.
\]
It defines the two light-like vectors $P_a$, $P_b$ that represent
the momenta of the parent partons prior to the gluon emission:
\begin{subequations}
\beeq 
p^{(a)}_a &=& zP_a + (1-z)yP_b - k_t , \\
q &=& (1-z)P_a + zyP_b + k_t,
\eeeq
\end{subequations}
where $k_t$ is ortogonal to $P_a$ and $P_b$.  
The on-mass-shell condition $p_a^2=0$ gives
\begin{equation}
 (P_a+yP_b-q)^2=0 \quad\Longrightarrow\quad
 y=\frac{(qP_a)}{(P_a-q,P_b)}.
\end{equation}
The light-cone fraction $z$ of the parent momentum $P_a$
carried by the final quark $p_a$ is given by the expression
\[
  1-z = \frac{(qP_a)}{(P_aP_b)}.
\]
In the collinear limit $y\to0$. In the soft limit both $y\to0$ and $z\to1$.

We now recast the phase space element in terms of the ``parent momenta'':
\begin{equation}
\begin{split}
 d\Phi_{ab}(q) &= (dp_a) (dp_b) (dq) (2\pi)^4\delta^{(4)}(p_a+p_b+q-Q)
\\ & = (dP_a) (dP_b) (2\pi)^4\delta^{(4)}(P_a+P_b-Q) \cdot (dq) \times
J^{(a)}_{ab}(q) \equiv d\Phi^{(0)}_{ab} \cdot (dq)\, J^{(a)}_{ab}(q).
\end{split}
\end{equation}
The Jacobian of this transformation reads
\begin{equation}
 J^{(a)}_{ab}(q) = \left(1-\frac{(qP_a)}{(P_a-q,P_b)}\right) \left(1-
  \frac{(qP_b)}{(P_aP_b)}\right)^{-1} = \frac{1-y}{z}.
\end{equation}
The gluon radiation probability we represent as a sum of two
contributions:
\begin{equation}
\label{eq:splitting}
 d\Phi_{ab}(q)\cdot W_{ab}(q) \>\>\Rightarrow\>\>
d\Phi^{(0)}_{ab} \cdot (dq)\, \left[
 W^{(a)}_{ab}(q)\,J^{(a)}_{ab}(q) \>+\>
 W^{(b)}_{ab}(q)\,J^{(b)}_{ab}(q) \right].
\end{equation}
Finally, the evolution equation in the differential form becomes  
\begin{subequations}\label{eq:Gen-fun}
\begin{equation}
\begin{split}
& E\partial_E\,
\cG(P_a,P_b;E) = 
 \int\frac{d\Omega}{4\pi}
\ab(q_t^{ab})\,\vartheta(q_t^{ab}\!-\!Q_0)\\[2mm]
&\times \sum_{c=a,b}
J_{ab}^{(c)}(q)\hW_{ab}^{(c)}(q)
\Big[u(q)\,\cG(p_a^{(c)},q;E)\,\cG(q,p_b^{(c)};E)  
\>-\>\cG(P_a,P_b;E)\Big]_{E=\omega}. 
\end{split}
\end{equation}
In the graphic form it can be represented as
\begin{equation}\label{eq:Gen-fun2}
 \epsfig{file=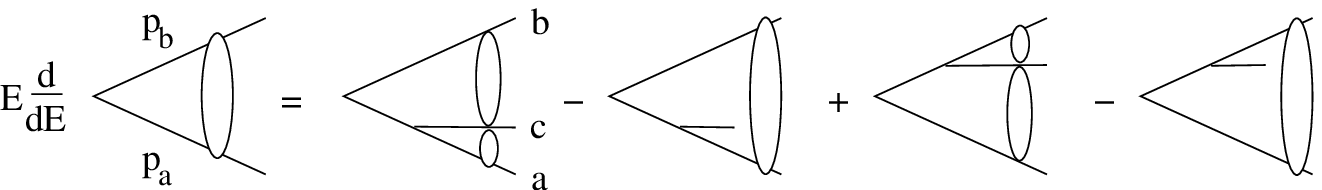}
\end{equation}
\end{subequations}
This evolution equation is the modification of \eqref{eq:softGF} that
includes recoil as described above. The gluon radiation function
$\widehat{W}_{ab}(q)=\omega^2W_{ab}(q)$ is split according to
\eqref{eq:splitting} into the two pieces which are collinear singular
when $q$ is parallel to $p_a$ or to $p_b$, correspondingly, see
\cite{CatSey}. 
The energy of the hardest real gluon $q_0\equiv\omega = E$ plays the r\^ole of the evolution parameter.

By introducing the Sudakov form factors this equation can be recast in
an integral form as it has been done in the soft case in
\eqref{eq:soft-GF-int}.  
\begin{equation}\label{eq:Gen-fun1}
\begin{split}
  \cG(P_a,P_b;E) \>=& \>\> 
S(P_a,P_b;E) + \int^E
\frac{d\omega}{\omega}\int \frac{d\Omega}{4\pi}\, \,u(q)\cdot
\frac{S(P_a,P_b;E)}{S(P_a,P_b;\omega)} \> \vartheta(q_t^{ab} - Q_0) \\ 
& \times \, \ab(q_t^{ab})  \sum_{c=a,b}
J_{ab}^{(c)}(q)\hW_{ab}^{(c)}(q) 
\cdot \cG(p_a^{(c)},q;\omega)\, \cG(q,p_b^{(c)};\omega) .
\end{split}
\end{equation}
This equation generalizes \eqref{eq:soft-GF-int} by taking into  account the energy--momentum recoil and again satisfies the normalization $\cG(P_A,P_b;E)=1$ for all $u(q)=1$ (cf.\ \eqref{eq:unity}).

An iterative solution of this integral
equation defines a Monte Carlo process for generating multi-parton
ensembles with account of collinear non-enhanced single-logarithmic
corrections due to large-angle soft gluon emission.

\medskip

Now we are going to check if the dipole recoil scheme described above
is consistent with known analytical results concerning the collinear
resummation.  The simplest observable of this type is an inclusive
energy distribution of the final-state quark.

\subsection{The non-singlet quark energy distribution}

The non-singlet quark fragmentation function is obtained by taking the
derivative of the generating functional $G_{ab}$ with respect to the
quark source $u(p_a)$ and setting all the remaining source functions
$u=1$.  This way we obtain the distribution in the momentum of the
quark $a$ accompanied by any number of gluons. If the quark energy is
taken large, $(1-x)\ll 1$, all radiated gluons are {\em soft}\/ and
the analysis simplifies significantly.  The problem becomes
essentially Abelian\footnote{Non-Abelian effects due to final state
  cascading of gluons reduce to making the effective coupling in the
  radiation probability of a primary gluon $k$ run with $k_\perp^2$.}
and is described by multiple independent radiation of soft gluons by
the quark.
 
\subsubsection{Collinear approximation}

We shall restrict ourselves to configurations in which all radiated
gluons have small emission angles with respect to the quark direction.
This --- quasi-collinear --- approximation is sufficient for the
analysis of the {\em anomalous dimension}\/ which accumulates
collinear (``mass'') singularities of the fragmentation function in
all orders and describes the scaling violation.

In this kinematics the measured variable $x$ is given by
\begin{equation} 
 x=1-\sum_iy_i\,,\qquad \om_i=y_i\,E \,,\qquad p_a\simeq x\,P_a.
\end{equation}
What matters is the total energy $1\!-\!x$ carried, on average, by an
ensemble of radiated soft gluons.

Looking for {\em collinear singular}\/ contributions --- the terms of
the type $\ab \ln(1-x)\cdot \ln Q_0$ --- we may omit the two last
terms on the r.h.s.\ of the equation \eqref{eq:Gen-fun2} that contain
the factor $W_{ab}^{(b)}(q)$ which is non-singular when the gluon
momentum $\vec{q}$ becomes collinear to $\vec{p}_a$.

One is then left with contributions involving only the singular
antenna piece $W_{ab}^{(a)}(q)$:
\begin{equation}\label{eq:neweq0}
 \epsfig{file=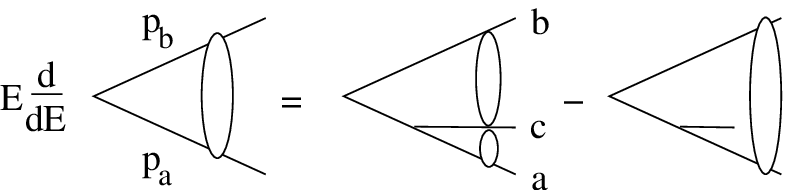}
\end{equation}
Since our observable is inclusive, we may use unitarity arguments to
simplify the equation.  Indeed, by the nature of the adopted recoil
strategy, production of gluons by the ``upper'' dipole $(bc)$ in
\eqref{eq:neweq0} does not affect the momentum of the quark $a$.
Therefore, since such emissions do not affect the measurement, they
are subject to real--virtual cancellation and can be neglected.

One is left to consider only gluons emitted in the lower blob
involving the parton $a$:
\begin{equation}\label{eq:neweq}
 \epsfig{file=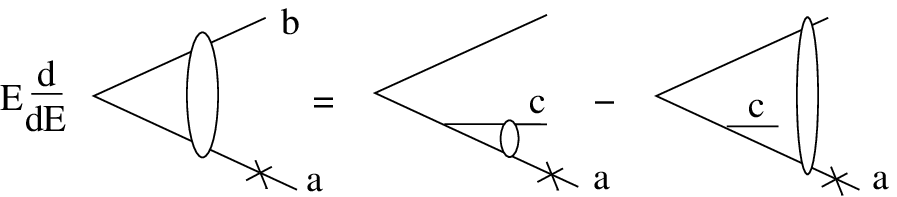}
\end{equation}
This equation generates multiple gluon emissions off the quark. 

By construction, successive emissions are ordered in gluon energies.
And here comes the crucial observation: as we have discussed above,
the emission angle of the gluon $c$ is essentially limited from above
by the aperture of the parent dipole, $\xi_{ab}$, see
\eqref{eq:ang-ord}.  Therefore, soft gluons generated by the evolution
equation \eqref{eq:neweq} turn out to be ordered, simultaneously, in
energies {\em and}\/ angles with respect to~the radiating quark.
Instead we know that the DGLA equation is obtaine from ordering only
in collinear variables, disregarding the relative energies of emitted
partons. The corresponding equation in the present dipole-formulation
looks as follows.

\subsubsection{Evolution equation with dipole recoil strategy}

Since the radiated gluons are soft, we use the distribution
$W_{ab}^{(a)}(q)$ defined in \eqref{eq:WAO} which describes the soft
part of the gluon radiation probability (splitting function).  We may
also approximately set $J^{(a)}_{ab}(q)=1$.

In going from the generating functional $\cG(P_a,P_b;E)$ in
\eqref{eq:Gen-fun} to the single particle distribution, we express the
latter as a function of the {\em angle}\/ between the partons
$\xi_{ab}$, the {\em energy scale}\/ $E\simeq E_a$ and the {\em energy
  fraction}\/ of the triggered quark: $D(\xi_{ab},E,x)$.

Performing integration over the azimuthal angle $\phi_{aq}$ 
(see \eqref{eq:ang-ord}) one deduces from \eqref{eq:Gen-fun}
\begin{subequations}\label{eq:badD}
\begin{equation}\label{eq:NSdistr-0}
\begin{split}
D(\xi;E;x)=\delta(1-x)+
&\int^1\frac{dy}{y}\int^{\xi}\frac{d\xi'}{2\xi'}\>\ab\,
\vartheta\left(yE\sqrt{\xi'}-Q_0\right)\\
&\times
\left[D\left(\xi',yE,\frac{x}{1-y}\right) - D\left(\xi',yE,x\right)\right],
\end{split}
\end{equation}
where $\xi=\xi_{ab}$ is the angular aperture of the original dipole
$(ab)$ while $\xi'=\xi_{ac}$ is the aperture of the secondary
$(ac)$--dipole, and for the sake of simplicity we have restricted ourselves to the soft part of the gluon radiation probability, $dy/y$. 
Equivalently one may write\footnote{The $y$ integration in \eqref{eq:NSdistr}
  is actually limited from above by $y=1-x$ since the distribution
  $D(\xi,Q,z)\propto \Theta(1-z)$} 
\begin{equation}\label{eq:NSdistr}
	\begin{split}
	D(\xi;E;x)& =\delta(1-x)\,S(E\sqrt{\xi})\\
	+&\int^1\frac{dy}{y}\int^{\xi}\frac{d\xi'}{2\xi'}\>\ab\,
	\vartheta\left(yE\sqrt{\xi'}-Q_0\right)
	\frac{S(E\sqrt{\xi})}{S(yE\sqrt{\xi})}
	\, D\left(\xi',yE,\frac{x}{1-y}\right),
	\end{split}
\end{equation}
\end{subequations}
with the Sudakov form factor given by the expression
\begin{equation}
\ln S(E\sqrt{\xi})=-\int^1 \frac{dy}{y} \int^{\xi}\frac{d\xi'}{2\xi'}
\ab\,\vartheta\left(yE\sqrt{\xi'}-Q_0\right).
\end{equation}
The distribution $D$ describing real gluon emission 
in the integrands of \eqref{eq:badD} is evaluated at $\xi'$ since a
gluon produced by the dipole $(ac)$ has an angle with respect to $a$ effectively
bounded from above by $\xi_{ac}$, see \eqref{eq:ang-ord}. 

It is worth noticing that going from the equation for the generating functional \eqref{eq:Gen-fun} to that for the inclusive quark spectrum, we have replaced the full angle aperture $\xi=\xi_{ab}$ by the opening angle $\xi'$ of the $(ac)$ dipole in the {\em virtual}\/ (subtraction) term as well. In so doing we followed the  preceding observation according to which virtual radiation of gluons constituting the ``wider'' dipole $(bc)$ had been already cancelled by corresponding real emissions, without affecting the energy of the triggered quark.   

Equations \eqref{eq:badD} correctly satisfy the sum rule for the first moment:
integrating over $x$ one obtains $1$ for the number of quarks $a$ (to
see this one uses \eqref{eq:no-emission}).  
However, they are in conflict with the Poisson nature of the multiple
soft gluon radiation and contradict the collinear resummation result 
leading to the DGLAP equation. 

\subsubsection{Analysis of the equation} 

Introducing the transverse momentum scale variable $Q=E\sqrt{\xi}$ and the corresponding integration variable $q=\omega\sqrt{\xi'}$ we get
\begin{equation}\label{eq:inq}
	D(x,Q) = \delta(1-x)+
\int^1_{Q_0/Q}\frac{dy}{y}\int^{yQ}_{Q_0}\frac{dq}{q}\>\ab \,
\left[D\left(\frac{x}{1-y},q\right) - D\left(x,q\right)\right],
\end{equation}
or, in terms of the standard Mellin moment representation, 
\begin{equation}\label{eq:inqN}
	D_N(Q) = 1 +
\int^1_{Q_0/Q}\frac{dy}{y}\int^{yQ}_{Q_0}\frac{dq}{q}\>\ab \,
\Big[(1-y)^N-1\Big]\,  D_N\left(q\right).
\end{equation}
In the region of large $x$, such that $1-x\ll 1$ (but $1-x \gg Q_0/Q$), the essential Mellin moments are large, $N\sim (1-x)^{-1}$, and cut from below the $y$-integral:
\begin{equation}\label{eq:inqN2}
	D_N(Q) \>\simeq\> 1 -
\int^1_{1/N}\frac{dy}{y}\int^{yQ}_{Q_0}\frac{dq}{q}\>\ab \,  D_N\left(q\right).
\end{equation}
By iterating this equation one obtains an oscillating double logarithmic series in $\ln Q$ and $\ln N$, 
$$
  D_N^{(n)}(Q)\> \sim\> \frac{\left(-\ab\ln N \ln {Q}\right)^n}{(n!)^2} \>+\> \ldots.
$$
Here one combinatorial factor $1/n!$ comes from the kinematical ordering of gluon {\em energies}, and the second  $1/n!$ from the ordering of the {\em angles}\/ with respect to the quark of the successively radiated gluons.  
This series sums up into a {\em Bessel function}\/  instead of an {\em exponent}\/ in $\ln Q$ with the well known DGLAP ``anomalous dimension''
$$
  D_N^{(n)}(Q)\> \sim \> \frac{\left(-\ab\ln N \ln { Q}\right)^n}{n!}, \quad D_N(Q) \propto Q^{-\ab\ln N},
$$
where we have neglected for the sake of simplicity the running of the coupling.  

In general (whether the coupling runs or not), the anomalous dimension $\gamma_N$ is defined as
$$
   \gamma_N(\ab(Q))\> =\> \frac{d}{d\ln Q} \ln D_N(Q).
$$
For $\ab=\mbox{const}$ this gives 
$$
  D_N(Q)= \left(\frac{Q}{Q_0}\right)^{\gamma_N(\ab)}.
$$
Substituting this ansatz  into \eqref{eq:inqN2}  and evaluating the logarithmic derivative, we obtain an equation for the new anomalous dimension:
\begin{equation}
\gamma_N(\ab)  \>\simeq\> -\ab \int_{1/N}^1 dy \, y^{\gamma_N-1} = -\ab \frac{1-\exp(-\gamma_N\ln N)}{\gamma_N}.
\end{equation}
Perturbatively, when $\ab\ln^2N\ll 1$, we have the correct one loop expression
\begin{equation}
   \gamma_N(\ab) \>\simeq\> -\ab\,\ln N + \cO{\ab^2\ln^3 N},
\end{equation}
while in higher loops it starts to deviate significantly from the DGLAP answer. 
In general, introducing the variable $w$
$$ \gamma_N= -\frac{w}{\ln N},
$$
the anomalous dimension is given by the solution of the equation
\begin{equation}
\ab \ln^2N = \frac{w^2}{e^w-1}.
\end{equation}
The r.h.s.\ hits the maximum at certain $w=w_0$, so that  
in the region of large moments $N$, starting from $\ab\ln^2N=\mbox{const}$, the anomalous dimension becomes complex valued and the distribution starts to oscillate.  

The origin of the failure of the energy ordering strategy 
can be understood already at the level of two emitted gluons.

\subsubsection{Two gluon emission}

Consider the emission of two soft gluons $p_1,p_2$ off the parton $q$
of the $q\bar{q}$ dipole. The antenna functions that potentially
contribute in the collinear limit (recall that we keep all gluon
angles with respect to the quark $a=q$ to be small) are displayed 
here: 
\beq\label{eq:3figs}
 \epsfig{file=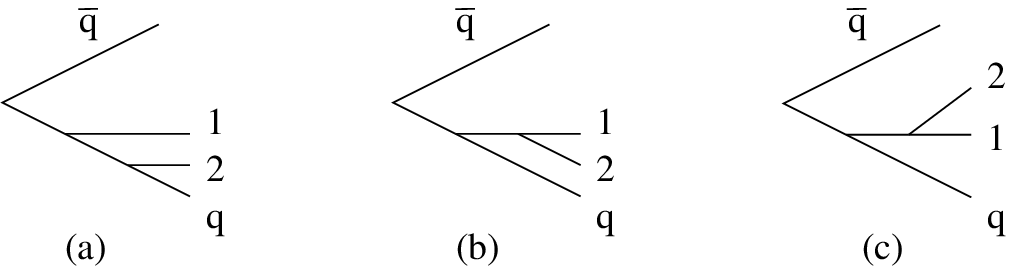,width=0.7\textwidth}
\eeq
The first two graphs correspond to the splitting of the dipole $(a1)$:
\[
 W_{a1}^{(a)}(2) \to \vartheta(\xi_{a1}-\xi_{a2})
\qquad {\rm and} \qquad
W_{a1}^{(1)}(2)\to \vartheta(\xi_{a1}-\xi_{12}),
\]
while the third one, 
\[
  W_{1\bar{q}}^{(1)}(2)\to \vartheta(\xi_{1\bar{q}}-\xi_{12}),
\]
is the relevant part of the large-aperture dipole $(1\bar{q})$.  Due
to the {\it local}\/ recoil prescription used, only the contribution
$(a)$ affects the momentum of the quark $q$.  In the two remaining
ones, (b) and (c), the gluon 2 borrows its energy--momentum from the
gluon 1 and does not produce any quark recoil. Therefore, these
contributions cancel against corresponding virtual corrections in the
inclusive quark measurement.
In conclusion, within the adopted recoil strategy, only the graph (a) should be kept, and we obtain the following phase space for the two-gluon emission:
\begin{equation}
\label{eq:orering2}
\xi_{a2} \ll \xi_{a1} \ll \xi_{a\bar{q}} 
\qquad \mbox{and} \qquad 
\omega_2 < \omega_1\ll E. 
\end{equation}
The first condition comes from the angular ordering in the graph
\eqref{eq:3figs}(a), and the second condition --- from the energy
ordering of successive emissions.

We know, however, that in order to obtain the DGLAP equation that
properly resums collinear singular contributions, one needs to
assemble angular (or, transverse momentum) ordered emissions (the
first ordering), regardless to the order of gluon energies (the second
one). At the same time, the dipole logic is leading us to the {\em
  double-ordered}\/ gluon ensemble, according to \eqref{eq:orering2}.
As a result, instead of a simple Poisson distribution of soft
radiation ($1/n!$) we obtain something like a Bessel function
distribution ($1/(n!)^2$).

What is missing here is actually the coherence of QCD radiation. As
well known, a soft gluon $\omega_2$, with $\omega_2\ll\omega_1\ll
E_a$, could be emitted at large angles ($\xi_{a2}\gg\xi_{a1}$)
directly by the original parton $q+p_1\simeq q$.  In the language of
Feynman amplitudes, such radiation occurs as a coherent sum of the
graphs (b) and (c).

\bigskip

Another way to discuss this point is the following.
The dipole radiation pattern of soft gluon $2$ off the composite
antenna $(a1\bar{q})$ can be represented as follows:
\begin{subequations}\label{eq:original}
\begin{equation}\label{eq:original-a}
W_{a1}(2)+W_{1\bar{q}}(2) =  
\bigg[W_{a1}(2)+W_{1\bar{q}}(2) - W_{a\bar{q}}(2) \bigg]
 + W_{a\bar{q}}(2). 
\end{equation}
Beyond the large-$N_c$ approximation, the latter term (corresponding
to emission of the gluon $2$ by the quark-antiquark dipole) acquires a
$N_c^2$-suppressed correction, $N_c\to N_c-1/N_c$:
\begin{equation}
  \frac{N_c}{2} \bigg(W_{a1}(2)+W_{1\bar{q}}(2)\bigg) \>
\Longrightarrow\> \frac{N_c}{2}\cdot \bigg[W_{a1}(2)+W_{1\bar{q}}(2) - 
W_{a\bar{q}}(2) \bigg]  + C_F\cdot W_{a\bar{q}}(2). 
\end{equation}
\end{subequations}
This is in line with the collinear factorization theorem. Indeed, the
factor in square brackets is {\em non-singular}\/ when $\vec{p}_2$
becomes parallel to the direction of the quark or antiquark momentum
so that these collinear enhancements are contained by the quark dipole
term $W_{a\bar{q}}(2)$ and must be proportional to the ``colour
charge'' (Casimir operator) $C_F$ of the quark.

The only collinear singularity of the first contribution on the
r.h.s.\ is that of $\vec{p}_2$ parallel to $\vec{p}_1$ which suggest
an interpretation of this term as independent emission of the gluon
$2$ by the previous generation (harder) gluon $1$.  The full two soft
gluon emission probability takes the form
\begin{equation}\label{eq:HER}
dw(1,2)\>\propto\> C_F\frac{N_c}{2}\cdot W_{a\bar{q}}(1)
\bigg[W_{a1}(2)+W_{1\bar{q}}(2) - W_{a\bar{q}}(2) \bigg]
 + C_F^2\cdot W_{a\bar{q}}(1)W_{a\bar{q}}(2). 
\end{equation}
In the large-$N_c$ limit, this is identical to the original dipole
expression.  At the same time, the representation \eqref{eq:HER}
suggests a different recoil strategy when it comes to incorporating
the energy--momentum conservation.

The first term in \eqref{eq:HER} describes cascade gluon
multiplication: $q\bar{q}\to 1$ followed by $1\to 2$. In the inclusive
quark distribution this is subject to the real--virtual cancellation.
Energy ordered configurations, $\omega_2\ll\omega_1$ cancel leaving no
trace; ``hard'' gluon splittings with $\omega_2\simeq\omega_1$ modify
{\em one gluon}\/ emission by giving rise to $\alpha_s(k_{1\perp}^2)$
for the radiation intensity of the primary gluon $1$ by the quark.

The second term describes independent radiation of two gluons off the
quark --- the second order term of the Poisson distribution of primary
gluons --- which is necessary to correctly recover the DGLAP quark
fragmentation function.

Equation \eqref{eq:HER} provides a probabilistic representation for the
two-gluon production in terms of a combination of independent and
cascade gluon emissions.  With account of hard parton splittings and
of the natural energy recoil prescription, it served as the base for
constructing the HERWIG parton event generator \cite{EAO,HERWIG}. The
corresponding probabilistic representation for production of three and
four energy ordered gluons was developed in \cite{DT} and described in
\cite{Basics}, which resulted in a formulation of the parton
multiplication scheme based on the exact angular ordering of
successive parton splittings.

\bigskip 

The difference between the two strategies becomes transparent. In the
dipole scheme, successive gluons borrowing the energy from the quark
were ordered in emission angles,
\[
 W^{(a)}_{a\bar{q}}(1)\, W^{(a)}_{a1}(2)\,, \qquad \xi_{2a}< \xi_{1a}<1, 
\]
while in \eqref{eq:HER} the angle of the softest gluon is not bounded,
and the two emissions are totally independent,
\[
 W^{(a)}_{a\bar{q}}(1)\, W^{(a)}_{a\bar{q}}(2)\,, \qquad 
\xi_{2a}<1, \>\> \xi_{1a}<1. 
\]
Let us see how this has happened, from the point of view of the
gluon--gluon splitting.

In the quasi-collinear configuration, $\xi_{a1}\ll 1$, the radiation
pattern of the softest gluon $2$ consists of a ``narrow'' and a
``wide-angle'' dipoles, $(a1)$ and $(1\bar{q})$.  When, within the
dipole recoil strategy, the ``wide-angle'' antenna $W_{1\bar{q}}(2)$
is split into the pieces collinear singular with respect to the
directions of the first generation gluon ($1$) and of the antiquark,
\[
 W_{1\bar{q}}(2) \>=\> W^{(1)}_{1\bar{q}}(2) + W^{(\bar{q})}_{1\bar{q}}(2), 
\]
the contribution $W^{(1)}_{1\bar{q}}(2)$ is looked upon as radiation
of the gluon 2 by the gluon 1, whichever the kinematical configuration
of the gluon 1 with respect to its parent quark (dipole
factorization).  At the same time, within the HERWIG logic (collinear
factorization), the gluon--gluon multiplication occurs only in a
restricted angular aperture, $\xi_{21}< \xi_{1a}$.  Indeed, in the
kinematical region $\xi_{1a}\ll \xi_{21}\ll 1$ the two last terms in
the combination
\[
 \bigg[W_{a1}(2) + W_{1\bar{q}}(2) - W_{a\bar{q}}(2) \bigg]
 \]
on the r.h.s.\ of \eqref{eq:original} cancel: 
\[
   \lim_{\xi_{1a}\to 0} \xi_{1a}^2\cdot 
\bigg[ W_{1\bar{q}}(2) - W_{a\bar{q}}(2) \bigg] = 
\cO{\frac{\xi_{1a}}{\xi_{12}}} \to 0
\]    
(while the first antenna term, $W_{a1}(2)$, is restricted in angle on
its own).

Physically, the gluon 2 radiated off the gluon 1 in the ``wide''
dipole \eqref{eq:3figs}(c) at {\em relatively large}\/ angles,
$\xi_{21} > \xi_{1a}$,
is produced by the grandparent parton $a$ (quark).  Correspondingly,
it must borrow energy from the quark rather than {\em locally}\/ from
the dipole $(1\bar{q})$ to which it belongs from the point of view of
the colour connection topology.

Reformulating this way the local dipole recoil prescription modifies
the energy ordered evolution equation.  The modification reduces to
substituting the full dipole opening angle $\xi=\xi_{ab}$ for the running
angle parameter $\xi'$ in the argument of the distribution $D$ on the
r.h.s.\ of \eqref{eq:badD}. As demonstrated in the Appendix, this way one
recovers the approximate DGLAP equation for the quark fragmentation function.

\section{Conclusions}

Monte Carlo generation of QCD events is a quarter century old business,
based on the structure of resummation of {\em collinear enhanced}\/
Feynman diagram contributions. The probabilistic parton cascade
picture was first established for one-particle inclusive quantities
(DGLAP evolution equations for DIS structure functions and $e^+e^-$
fragmentation functions, \cite{DGLAP}),
generalised via ``jet-calculus'' \cite{KUV} and, with account of the
soft gluon coherence \cite{AO,BCM,EAO,Basics}, has laid the base for
probabilistic description of the internal structure of parton jets and
their ensembles.

The corresponding coherent branching Monte Carlo schemes (HERWIG, in particular) reproduce 
correctly both the leading double logarithmic effects (LL) in various QCD observables as well as subleading single logarithmic collinear-enhanced terms. 
One may recall as examples NLL corrections to the mean particle multiplicity \cite{B} and to the inclusive particle spectra from light parton \cite{A,C} and heavy quark initiated jets \cite{E}, to global event shapes in $e^+e^-$ annihilation \cite{F}, as well as many other quantities sensitive to parton cascades.   

{\em Collinear-non-enhanced}\/ (``large angle'') soft gluon radiation
provides significant NNLL corrections to global event characteristics
(e.g., mean particle multiplicity); it is also responsible for
inter-jet multiplicity and energy flows and determines the structure
of various {\em non-global}\/ observables
\cite{DasSal,ng-plus,DM_LargeAng}.  Effects of multiple soft gluon
radiation at large angles lie beyond the scope of the standard
(collinear) approach and must be treated order by order in
perturbation theory (while collinear enhanced contributions are
resummed in all orders).

An elegant expression \cite{BCM} for the multiple soft gluon
production probability \eqref{eq:BCM-ant} is valid for {\em arbitrary
angles}\/ and offers a possibility of improving the parton picture.
The structure of multi-gluon distribution \eqref{eq:BCM-ant} naturally
suggests an interpretation in terms of a chain of {\em colour
connected dipoles}.  By choosing an ``evolution time'' variable this
chain may be generated via a Markov process of successive dipole
splittings. An evolution time can in principle be chosen differently.
We have used energy ordering of gluons for this purpose. The
generating functional that we have constructed with the help of the
corresponding evolution equation allows one to calculate specific
effects due to multiple emission of soft gluons at large angles in the
large-$N_c$ approximation.

In order to construct a realistic Monte Carlo generator for
multi-parton ensembles it is imperative, however, to formulate an
adequate recoil prescription which would ensure energy--momentum
conservation at every successive step of the parton (dipole)
multiplication.

In the present paper we addressed the question, whether the ``dipole
factorization'' extends beyond the no-recoil approximation. In other
words, whether the splitting of a colour dipole into two can be
treated independently of the prehistory of the system, that is {\em
  locally}\/ in the evolution time (which is a necessary condition for
constructing a Markov process).

Having taken an inclusive energy distribution of the final quark as
the simplest example of a collinear sensitive observable, we have
shown that a naive implementation of the dipole recoil strategy
results in violating the collinear factorization. 

However, the existing Monte Carlo implementations of the dipole
picture  (for recent examples see for instance
\cite{dipoleMC0,dipoleMC1,dipoleMC2}) chose to order successive gluon
emissions in (dipole centre of mass) {\em transverse momenta}\/ rather
than in (laboratory frame) gluon energies. 
Such a construction apparently prevents the problem under
consideration from appearing in the {\em leading collinear approximation}, 
that is in all orders in the QCD coupling corresponding to the one-loop level in the anomalous dimension. 
 
Indeed, as we have seen above, the problem with the quark recoil
appears when a softer gluon (2) is emitted at a {\em large}\/ angle
than its predecessor (1). The inclusive quark energy distribution is
sensitive to the region where the two gluons have comparable energies.
Then, a softer one that carries a slightly smaller energy but is
radiated at a large angle, acquires a larger {\em transverse
  momentum}, and therefore, by construction, is looked upon as ``the
first'' by the $k_\perp$-ordered MC scheme.  Therefore, ``the second''
gluon in such a picture, that is the one with smaller transverse
momentum, is automatically having a smaller emission angle; in the
leading logarithmic approximation the coherence is respected.

The problem of an interplay between energy--momentum recoil and soft
gluon coherence discussed in this paper gets postponed but should be
kept in mind in view of attempts at constructing the next-to-leading
order QCD MC generators.

\section*{Acknowledgements}
We thank the organisers and participants of the CERN TH jet
phenomenology seminar for the opportunity to extensively discuss the
issues raised in the paper.  We are especially grateful to Walter
Giele, G\"osta Gustafson, David Kosower, Leif L\"onnblad, Zoltan Nagy,
Gavin Salam, Jim Samuelsson, Mike Seymour, Torbjorn Sj\"ostrand, Peter
Skands, Davidson Soper and Bryan Webber for illuminating discussions.

\appendix
\setcounter{equation}{0}

\section{Global recoil and correspondence with DGLAP}

We have no universal recipe under sleeve for extending the dipole
gluon multiplication picture beyond the no-recoil approximation.
However, for a simple example of the inclusive quark distribution
discussed in Section~\ref{Sec3} the situation is straightforward to cure.

As we saw from the previous discussion, the origin of the failure lay
in the fact that the softest gluon emitted at angles {\em larger}\/
than the angle of the previous branchings did not contribute to the
quark recoil.  So, it suffices, by brute force, to permit the gluons
of all generations to contribute to the quark recoil, irrespectively
to the value of their emission angles.

This amounts to replacing in the integrand of \eqref{eq:NSdistr} the
distribution $D$ evaluated at $\xi=\xi_{ac}$ with the distribution
evaluated at the full opening angle $\bar\xi=\xi_{ab}$.  
One then has the different equation,
\begin{equation}\label{eq:NSdistr-DGLAP}
\begin{split}
D(\bar\xi,E;x)=&\delta(1-x)\,S(E\sqrt{\bar\xi})\\
+&\int^1\frac{dy}{y}\int^{\bar\xi}\frac{d\xi}{\xi}\frac{\ab}{2}
\vartheta\left(yE\sqrt{\xi}-Q_0\right)
\frac{S(E\sqrt{\bar\xi})}{S(yE\sqrt{\bar\xi})}
\, D\left(\bar\xi,yE;\frac{x}{1-y}\right),
\end{split}
\end{equation}
which actually corresponds to the DGLAP evolution. 
To see this, we iterate equation \eqref{eq:NSdistr-DGLAP} to obtain the series
\begin{equation}
\label{eq:NS-5}
\begin{split}
&D(\bar\xi,E;x)=S(E\sqrt{\bar\xi})\\
&\times \left\{\delta(1-x)+\sum_{n=1}^\infty
\prod_i\left(\int^1\frac{dy_i}{y_i}\int^{\bar\xi}\frac{d\xi_i}{\xi_i}\,
\frac{\ab}{2}\,\theta(y_iE\sqrt{\xi_i}-Q_0)\right)\cdot 
\delta\left(1-\frac{x}{1-\sum y_i}\right)
\Theta_{\rm en-ord.}\right\},
\end{split}
\end{equation}
where 
\begin{equation}
\Theta_{\rm en-ord}\>=\>\Theta(y_n<\cdots<y_1<1)
\end{equation}
is the product of with theta-functions which ensures the energy
ordering of gluons.  Now, due to the symmetry of the multiple integral
\eqref{eq:NS-5} with respect to energy and angular variables, we can
replace the {\em energy ordering}\/ with the {\em angular ordering}\/:
\begin{equation}\label{eq:angord}
\Theta_{\rm en-ord}\>\>\Longrightarrow\>\> \Theta_{\rm ang-ord}=
\Theta(\xi_n<\cdots<\xi_1<\bar\xi).
\end{equation}
Now the gluons are no longer ordered in energies (the gluon $q_i$ does
not need to be softer then $q_{i-1}$) but in the emission angle with
respect to the quark. In the angular ordered form, a soft gluon
radiated at a relatively large angle can be said to be emitted {\em
  before}\/ any other (harder or softer) gluons that move in a
collinear bunch around the quark. This is in accord with the colour
coherence according to which a large-angle soft gluon ``sees'' only
the total colour charge of a collinear group of partons.

To derive the DGLAP equation we replace
\begin{subequations}\label{eq:ykinem}
\begin{equation}
\frac{d\om_i}{\om_i}=\frac{dy_i}{y_i}=\frac{dz_i}{1-z_i}\,,\qquad
{1-\sum y_i}=z_1\cdots z_n 
\end{equation}
where now $1-z_i$ is the local fraction of energy taken away by the
gluon $q_i$ from the parent parton (quark), while $y_i$ is the gluon
energy fraction with respect to $E$, the energy of the primary parton
$P_a$:
\begin{equation}
y_i=(1-z_i)z_{i-1}\cdots z_1\,.
\end{equation}
\end{subequations}
We finally obtain the evolution equation
\begin{equation}
\label{eq:NS-8}
\begin{split}
D(Q,x)&=\delta(1-x)S(Q)
 + \int_{Q_0^2}^{Q^2}\frac{dq_t^2}{q_t^2}\int_0^{1-\frac{Q_0}{q_t}}
dz\,\frac{\ab(q_t)}{2(1-z)}\,\cdot \frac{S(Q)}
{S(q_t)}\,D\left(q_t,\frac{x}{z}\right)\\
&=\delta(1-x) + 
\int_{Q_0^2}^{Q^2}\frac{dq_t^2}{q_t^2}\int_0^{1-\frac{Q_0}{q_t}}
dz\,\frac{\ab(q_t)}{2(1-z)}\,\cdot \left(D\left(q_t,\frac{x}{z}\right)
-D(q_t,x)\right)
\end{split}
\end{equation}
with $Q=E\sqrt{\xi}$ and $q_t=yE\sqrt{\xi}$. The upper boundary of the
$z$-integration, $1-Q_0/q_t$, can be replaced by 1 in the logarithmic
collinear approximation, which results in the integral equation that
correctly gives (the soft piece of) the DGLAP anomalous dimension.
\bigskip

Finally, let us mention another obstacle that one faces when trying to
construct the full realistic Monte Carlo scheme starting from the
energy ordered gluon ensembles.

In order to match the full DGLAP anomalous dimension, one has to
include the ``hard part'' of the splitting function by adding to the
distributions $\widehat{W}_{ab}^{(c)}(q)$ the pieces vanishing at
$\omega= 0$.  In the equation \eqref{eq:NS-8} this calls for the
replacement
\begin{equation}
\frac{dy_i}{y_i}=\frac{dz_i}{1-z_i}
\quad \Longrightarrow \quad dz_i\left(\frac{1}{1-z_i}-\half(1+z_i)\right). 
\end{equation}
However, while we could cast the leading soft distribution in
\eqref{eq:NSdistr-DGLAP} in terms of the local energy fraction,
$dy_i/y_i=dz_i/(1-z_i)$, the finite term $(1+z_i)\,dz_i$ cannot be
expressed {\em locally}, via the single variable $y_i$, see
\eqref{eq:ykinem}.

\end{document}